# LR(1) Parser Generation System:
# LR(1) Error Recovery, Oracles, and Generic Tokens

Arthur Sorkin
Web Oasis, Inc, 940 N Barkley, Mesa, AZ 85203, USA
Email: lr1@web-oasis.com

Peter Donovan
Adobe Systems Inc., 345 Park Avenue, San Jose, CA 95110

**Abstract**

The LR(1) Parser Generation System generates full LR(1) parsers that are comparable in speed and size to those generated by LALR(1) parser generators, such as yacc [5]. In addition to the inherent advantages of full LR(1) parsing, it contains a number of novel features. This paper discusses three of them in detail: an LR(1) grammar specified automatic error recovery algorithm, oracles, and generic tokens.

The error recovery algorithm depends on the fact that full LR(1) parse tables preserve context. Oracles are pieces of code that are defined in a grammar and that are executed between the scanner and parser. They are used to resolve token ambiguities, including semantic ones. Generic tokens are used to replace syntactically identical tokens with a single token, which is, in effect, a variable representing a set of tokens.

## 1. Introduction and Background

The LR(1) Parser Generation System generates full LR(1) parsers that are comparable in speed and size to those generated by LALR(1) parser generators, such as yacc [5] . Full LR(1) parsers have an inherent advantage over LALR parsers in that, *inter alia*, every deterministic context-free language can be recognized by some LR(1) parser. In addition, the LR(1) Parser Generation System contains a number of novel features. This paper discusses three of them in detail: an LR(1) grammar specified automatic error recovery algorithm, oracles, and generic tokens.

The error recovery algorithm depends on the fact that full LR(1) parse tables preserve context. Oracles are pieces of code that are defined in a grammar and that are executed between the scanner and parser. They are used to resolve token ambiguities, including semantic ones. Generic tokens[1] are used to replace syntactically identical tokens with a single token, which is, in effect, a variable representing a set of tokens.

LR(1) parsing was originally defined by Knuth [6]. SLR(1) and LALR(1) parsing were first defined by DeRemer [1], [2], [3]. SLR(1) parsers are constructed from the LR(0) CFSM[2] with "simple" lookahead sets added for inadequate[3] states. LALR(1) is the largest set of parsers that can be constructed from the LR(0) CFSM by adding lookahead sets. LALR(1) parsers are a proper subset of full LR(1) parsers.

Knuth's definition of LR(1) parsing carries along right context as part of a state. However, this requires an exponential number of states, which made full LR(1) parsers impractical. Pager solved this problem by showing that compatible states can be merged on the fly during LR(1) parser construction [10]; the resulting parsers are similar in size and speed to LALR(1) parsers. The first well known Pager LR(1) implementation was done by Shannon and Wetherell at Lawrence Livermore National Laboratory (LLNL) [11]. O'Hair, also at LLNL, updated and improved Shannon and Wetherell's implementation [9].

The LR(1) Parser Generation System (called "LR" below for brevity) consists of an LR(1) grammar analyzer, which contains an original optimized implementation of Pager's algorithm[4], a parser file constructor that takes the generic outputs from the grammar analyzer and injects them into a skeleton parser file[5], and a set of support routines for the skeleton parser. Splitting grammar analysis from parser construction is more flexible than tying analysis output to a particular parser skeleton or language.

Generic tokens (with subtokens and dynamic operator precedence), and the grammar specified automatic error recovery algorithm (discussed below) were originally implemented by the first author in a copy of the UCLA SLR(1) generator system [7], [8] written by Martin. While generic tokens worked correctly, it was discovered that the error recovery algorithm could fail in some cases. LR(1) parsing was required to fix the problem because its states always retain correct right and left context. LR, as described in this paper, was the eventual result. Oracles were implemented in later LR versions.

LR has been used in a number of commercial projects dating from its first version in 1987, and has added features over the years. The version of LR described in this paper is 5.x, which has been released as open source[6].

## 2. LR(1) Error Recovery

**EOF** is a built-in token that is returned by the scanner when an actual end of input is encountered. **ERROR** is a built-in token that is used to mark a place in the right hand side of a production where an error recovery action can happen. **ERROR** is not returned by the scanner and only appears in the grammar, and, as a result, in the parser.

---

[1] A token returned by the scanner corresponds to a terminal symbol in the grammar. For historical reasons the terms "terminal symbol" (or just "terminal"), and "token" are used interchangeably in this paper when describing LR. Terminals in LR are either reserved or plain. Reserved terminals are literals, e.g., 'goto', that specify the actual string the scanner is to match. Plain terminals, e.g., id, are other things the scanner matches.

[2] Characteristic Finite State Machine

[3] Having a read-reduce or reduce-reduce conflict.

[4] The optimized Pager LR(1) implementation will be discussed in a separate paper.

[5] Separate grammar analysis and parser construction is found in the LLNL Pager LR(1) implementation.

[6] http://sourceforge.net/projects/lr1/files/

A simple production (i.e., one with no alternatives) that has **ERROR** on its right hand side[7] is called an error production[8]. Error productions are commonly inserted to preserve enough information to prevent cascades of errors. Excessive use of error productions can cause an otherwise unambiguous grammar to become ambiguous (on **ERROR**). The LR error recovery algorithm requires that there be no ambiguous states caused by **ERROR**.

When the parser encounters an error in some state, it first, starting from that state, performs by normal parser actions all reductions, if any, which have **ERROR** in their right context set. This sequence of reductions is guaranteed to eventually result in the parser being in a state that has no reduction with **ERROR** in its right context set[9].

When such reductions, if any, have been done, the parser then tries to recover according to the following steps. It scans down the stack looking for a recoverable state[10]. A state is recoverable if that state has a read transition on **ERROR**, and the current scanner token is in the FIRST(1)[11] set of the successor state of the read transition on **ERROR**. In an LR(1) parser, FIRST(1) of a state is exactly the union of the tokens in the read transitions of the state and the tokens in the right context sets of the reductions in the state.

If there is no recoverable state on the stack, the parser causes the scanner to fetch the next token, which becomes the new current token, and then it looks for a recoverable state on the stack again. These two steps are repeated until a recoverable state is found. The parser will terminate in error recovery if it tries to discard **EOF** and fetch a new token.

When a recoverable state is found, recovery occurs by the parser taking the read transition on **ERROR** (as if **ERROR** had been returned by the scanner) and normal parsing resumes in the successor state. Since at the point of recovery, the current scanner token is in the FIRST(1) set of the successor state, parsing in guaranteed to resume without the current scanner token causing a new error.

Because scanning for a recoverable state takes place before getting a new token, it is possible in some cases to recover from an error without losing any information. Though it does neither, the recovery algorithm can behave as if it can insert a missing token or delete extra ones.

The following is an example production (with alternatives) that recognizes a comma separated id list:

```
<idlist> : <idlist> ',' id
     | <idlist> ERROR id
     | id  ;
```

The first example error listing below shows recovery with a missing comma. '**f**' triggers an error, and recovery is on '**f**'.

```
###     2 | int e f,g;
###             ^
#E  "tests/test4", line 2/7: syntax error
### Saw token: id='f'
### expected:  ; ,
###
###     2 | int e f,g;
###             ^
### Resuming parse with token: id='f'
```

In the following example production, a statement can be, among other things, an error terminated by a semicolon.

<st> : <decl-stmt> ';'
    | <c-stmt>
    | ';'
    | ERROR ';' ;

The second example error listing shows that an extra '**=**' triggers an error, and recovery is on the '**;**'.

```
###     5 |     a = =  b+c;
###                 ^
#E  "tests/test5.c", line 5/10: syntax error
### Saw token: =
### expected:  ( sizeof constant ++ -- & * + - id string_literal ! ~
###
###     5 |     a = =  b+c;
###                    ^
### Resuming parse with token: ;
```

It is customary for an LR input file to have an error production that specifies that the entire program can be an error. For example,

```
<user goal> : ERROR ;
```

where <user goal> is the left hand side of the first production[12]. If there is no other applicable recovery, this allows a recovery on EOF rather than just terminating the parse.

## 3. Oracles

An oracle as defined in LR is a fragment of code potentially executed between the scanner and parser that answers the question, "If the scanner returned token X, and the parser is looking for token Y in the current state, should Y replace X as the current scanner token?" The oracle's fragment of code is only executed if the current scanner token is X and the token Y is in the FIRST(1) set of the current parser state. If the oracle code fragment is executed, it decides if Y replaces X or not[13]. The parser then processes the current state with the resulting token.

This kind of oracle applies to all states and could be termed a global oracle. It is possible to imagine oracles that only apply in certain contexts, e.g., in particular states or sets of states, but context-specific oracles have not proven necessary in practice.

---

[7] A production with right hand side alternatives is actually shorthand for separate simple productions with identical left hand sides.

[8] Error productions are found in yacc [5], but its error recovery algorithm, while, therefore, having some similarities to LR's, works quite differently.

[9] The resulting state may or may not have ERROR as a read transition.

[10] Normal LR parser operation pushes states on the parse stack.

[11] FIRST(k) of a state in a DPDA is the set of k-length prefixes of token strings accepted starting in that state.

[12] LR has a built in 0th production because <user goal> requires context.:
`<GOAL> : EOF <user goal> EOF ;`

[13] In LR's oracle implementation, the input string seen by the scanner is immutable. So, some combinations of X and Y will be flagged as errors. For example, '**\***' cannot be changed to '**+**'.

Oracles are useful for resolving language ambiguities. For example, an identifier in C may, by virtue of a **typedef**, also be a type name. The scanner normally doesn't know which, so has to always return one or the other but not both. Let us assume that the scanner always returns token **id** with corresponding input text string contained in **Pval.text**.

The following oracle handles the ambiguity. The oracle code checks the symbol table to see if the **id** has been declared to be a **typedef**, and, if so, sets the special variable **oracle** to **TRUE**, which means the scanner token **id** gets changed to **TYPENAME**[14]. Otherwise, **oracle** is set to **FALSE**, and the current scanner token is not changed.

```
%oracle id: TYPENAME {
    if(is_typedef(Pval.text)) oracle = TRUE;
    else oracle = FALSE;  } ;
```

The following two C statements illustrate the operation of the above oracle for the **id** vs. **TYPENAME** token ambiguity. If we assume that the typedef is entered into the symbol table before the second statement is scanned, then the diagnostic outputs below show the token stream for each statement and the operation of the oracle above.

```
typedef unsigned int foo;
Token: typedef  [1:8]
Token: unsigned  [1:17]
Token: int  [1:21]
Token: id = 'foo' [1:25]
In ask_oracle with state 12 and token id
id not changed
Token: ;  [1:26]

foo b;
Token: id = 'foo' [3:4]
In ask_oracle with state 57 and token id
Token: id changed to Token: TYPENAME
Token: id = 'b' [3:6]
In ask_oracle with state 25 and token id
id not changed
Token: ;  [3:7]
```

Oracles are also useful in resolving symbol ambiguities between tokens and subtokens, which will be discussed in section 4.1.

## 4. Generic Tokens

Plain tokens in LR can be generic. A generic token represents a set of reserved terminals that are all handled the same way in the grammar. The use of generic tokens simplifies the grammar, reducing parse table size and often speeding up parsing by reducing parse tree height.

A subtoken is represented in an LR input file as a (generic) plain terminal, '.', and a reserved terminal. It represents a specific instance of a generic token's reserved token set. Both the dotted form and, when the generic token is known, a reserved token instance can be referred to as a subtoken[15]. For example,

**dualop.'*'** is a subtoken with **dualop** the generic token and **'*'** the reserved token instance of **dualop**, but in the context of **dualop,'*'** can also be called a subtoken. LR automatically notes which plain terminals are generic and which subtokens they have[16].

Since generic tokens in productions are, in effect, variables, the subtoken used needs to be known by the parser at runtime. This requires a generic token specification mechanism when the grammar is analyzed. 'The special symbols **%use** and **%ref** are used on the right hand side of productions to select the following token, which must be generic[17]. The use of **%use** and **%ref** will be discussed further below.

Subtokens, as opposed to generic tokens, cannot appear in productions. Subtokens, when they appear, are used in other kinds of rules, such as oracles, (node) maps, or precedence (all discussed below).

In the following discussion, we will assume that **dualop** is a generic token with subtokens **'+'**, **'-'**, **'*'**, and **'&'**, which are C operators that can be either unary or binary.

### 4.1 Generic Tokens and Oracles

Oracles are useful for resolving symbol ambiguities between plain tokens and instances of generic tokens, where both use the same symbol[18].

For example, in C, **'*'** can be an expression operator, or it can be used as part of a declaration (in which case, it is not an operator. The following productions show the use of **dualop** in C expressions and the reserved terminal token **'*'** in C declarations (**=>**, **%use**, and **%map** are discussed below).

```
<sexp> : <sexp> %use dualop <sexp>
    => %map infix ;

<pointer> : '*' <type-qual-list>  => n_pointer ;
```

In the following example oracle, if the scanner always returns subtoken **dualop.'*'** for **'*'** found in the input stream, then **dualop.'*'** is always replaced by token **'*'** in states with token **'*'** in FIRST(1).

```
%oracle dualop.'*': '*'   %{  oracle = TRUE; %} ;
```

The following two C statements illustrate the operation of the above oracle for the **dualop.'*'** vs. **'*'** symbol ambiguity. The associated diagnostic output shows the token stream for each statement and the operation of the oracle above.

```
int a, *b;

Token: int  [3:4]
Token: id = 'a' [3:6]
In ask_oracle with state 12 and token id
id not changed
```

---

is generic.

[16] In case a token has no subtokens appearing in an input file, LR has a special rule that is used to mark a plain terminal as generic and to define its subtokens.

[17] The implementation actually records the stack offset from the top of the stack where the subtoken will be stored at runtime.

[18] As above, the scanner input is immutable, e.g, changing **dualop.'+'** to **'*'** or to **dualop.'*'** is not allowed.

[14] So an ordinary token, e.g., **TYPENAME**, may appear in the grammar but never be returned by the scanner.

[15] In LR's implementation, the token data structure has a subtoken number field associated with it that is only meaningful if the token

```
Token: ,  [3:7]
Token: dualop Subtoken: *   [3:9]
In ask_oracle with state 85 and token dualop
Token: dualop changed to Token: *
Token: id = 'b' [3:10]
In ask_oracle with state 3 and token id
id not changed
Token: ;  [3:11]

a += *b;

Token: id = 'a' [5:2]
In ask_oracle with state 190 and token id
id not changed
Token: asop Subtoken: +=   [5:5]
In ask_oracle with state 106 and token asop
asop not changed
Token: dualop Subtoken: *   [5:7]
In ask_oracle with state 199 and token dualop
dualop not changed
Token: id = 'b' [5:8]
In ask_oracle with state 105 and token id
id not changed
Token: ;  [5:9]
```

*4.2 Generic Tokens and Tree Building Actions*

LR has tree building actions. Syntactically, a tree action[19] in LR is the symbol => followed by a (node) name[20]. For example,

`<expr> : <expr> '+' <arithexpr> => n_plus ;`

Conceptually, when the parser does a reduction, its tree action (if any) constructs a tree making the specified node the root node over any trees associated (via the stack) with the right hand side symbols (not every right hand side symbol will have an associated tree). After the reduction, the resulting tree is associated (via the stack) with the left hand side non-terminal[21]. LR uses a tree form in the parser where a parent node points to the leftmost element of an array of sibling nodes.

Generic tokens present a problem for tree building since they represent a set of subtokens. A normal tree building action with single node name will not suffice for a generic token, e.g., replacing '+' with **dualop** in the above production, there is no single node name that will be correct in all cases.

Map rules are used to specify node correspondence for generic tokens. Lexically, a map rule consists of a map name and a list of subtoken tree actions, which specify the node name used for the particular subtoken. The same subtoken can appear in more than one map and can specify a different node in different maps.

For example, both maps **infix** and **prefix** below contain **dualop.'*'** but map **prefix** specifies node **n_indirect** and map **infix** specifies node **n_mul**

```
%map infix : dualop.'&' => n_and
    | dualop.'+' => n_plus
    | dualop.'-' => n_minus
    | dualop.'*' => n_mul  ;

%map prefix : dualop.'&' => n_addr
    | dualop.'-' => n_uminus
    | dualop.'*' => n_indirect ;
```

Map names, which have no special meaning, are used in tree building actions in place of individual node names. At parser runtime, the subtoken instance of the generic token selected by **%use** or **%ref** determines the node name to be used based on the named map. **%use** and **%ref** are identical for tree building, but **%use** also specifies the use of dynamic precedence (discussed below in 4,3).

The following productions use the maps **infix** and **prefix** defined above for tree building (**%prec** will be discussed below in 4.3).

```
<sexp> : <sexp> %use dualop <sexp>
    => %map infix ;

<preexp> :  %ref dualop <castexp>
    %prec unop.'~' => %map prefix ;
```

*4.3 Generic Tokens and Precedence*

Precedence rules are used to assign both a precedence level and associativity to operator tokens[22]. Precedence allows grammar ambiguities involving operators to be resolved. This allows a simplified operator hierarchy in the grammar.

Precedence relationships for ordinary (non-generic) tokens are static and can be determined from the current scanner token and the production being reduced. However, since subtokens cannot appear in grammar productions, and since generic tokens are runtime variables, their precedence *must* be handled dynamically using the parse stack[23].

In LR the special tokens **%right**, **%left**, and **%noassoc** specify right associative, left associative, and non associative respectively. **%prec** is used to specify a precedence that is the same as another token's. However, LR allows subtokens in precedence rules as well as ordinary tokens.

The following rules define precedence for C operators using generic tokens **asop**, **op**, **dualop**, **unop**, and **doubleop**. **dualop's** subtokens are operators that can be either unary or binary, as are **doubleop's**.

---

[19] Tree building actions are found in MetaWare's TWS [4], but it doesn't support tree building for vocabulary actions or generic tokens.

[20] There is also a more complex n-tuple form where some flags and a function to be called can be specified in addition to the node name. The n-tuple form will be ignored in this discussion as it adds no additional complications for generic tokens.

[21] Nodes can also be associated with symbols via vocabulary action rules (also found in the LLNL Pager LR(1) implementation). A vocabulary action occurs when a token (ordinary or generic) is consumed by a read transition. We will ignore vocabulary actions in this discussion as well.

[22] Precedence rules are found in yacc, but yacc doesn't support precedence for subtokens.

[23] Actually, once the dynamic precedence mechanism exists, all token precedence can be handled dynamically.

```
%right : asop.'=' asop.'~=' asop.'*=' asop.'/='
asop.'%=' asop.'+=' asop.'-=' asop.'<<='
asop.'>>=' asop.'&=' asop.'^=' asop.'|=' ;
%left : op.'||' ;
%left : op.'&&' ;
%left : op.'|' ;
%left : op.'^' ;
%left : dualop.'&' ;
%left : op.'==' op.'!=' ;
%left : op.'>=' op.'<=' op.'<' op.'>' ;
%left : op.'>>' op.'<<' ;
%left : dualop.'+' dualop.'-' ;
%left : dualop.'*' op.'/' op.'%' ;
%left : unop.'~' unop.'!' ;
%left : doubleop.'++' doubleop.'—' ;
```

In a production, **%use** tells which token or subtoken to use for dynamic precedence. **%prec** specifies the use of another token or subtoken's precedence level. In the first production, below, **%prec** specifies that the precedence of **unop.'~'** should be used no matter the subtoken of **dualop**. In the second production, **%use** specifies the use of the dynamic precedence associated with **dualop's** subtoken.

```
<preexp> : %ref dualop <castexp>
    %prec unop.'~' => %map prefix ;

<sexp> : <sexp> %use dualop <sexp>
    => %map infix ;
```